\documentstyle[epsf,a4,12pt]{article}
\def\be{\begin{equation}}
\def\ee{\end{equation}}
\def\bq{\begin{eqnarray}}
\def\eq{\end{eqnarray}}
\def\ra{\rightarrow}
\begin{document}
\thispagestyle{empty}
\setcounter{page}{0}
\setcounter{page}{0}
\begin{flushright}
MPI-PhT/94-63 \\
LMU 16/94\\
September 1994
\end{flushright}
\vspace*{\fill}
\begin{center}
{\Large\bf Heavy Meson Form Factors, Couplings and Exclusive
Decays in QCD}$^*$\\
\vspace{2em}
\large
A. Khodjamirian$^{a,b,1}$ and R. R\"uckl$^{a,c,2}$\\
\vspace{2em}
{$^a$ \small Sektion Physik der Universit\"at
M\"unchen, D-80333 M\"unchen, Germany }\\
{$^b$ \small Yerevan Physics Institute, 375036 Yerevan, Armenia } \\
{$^c$ \small Max-Planck-Institut f\"ur Physik, Werner-Heisenberg-Institut,
D-80805 M\"unchen, Germany}\\

\end{center}
\vspace*{\fill}

\begin{abstract}
We discuss the form factors of the
heavy-to-light transitions $B \ra \pi$ and $D\ra \pi$,
the $B^*B \pi$ and $D^*D \pi$ coupling constants,
and the nonfactorizable amplitude of the decay $ B \ra J/\psi K$
in the framework of QCD sum rules.
\end{abstract}

\vspace*{\fill}

\begin{flushleft}
\noindent$^1$ Alexander von Humboldt Fellow\\
\noindent$^2$ supported in part by the German Federal Ministry for Research and
Technology (BMFT) under contract No. 05 6MU93P\\
\noindent$^*$
{\it presented by R. R\"uckl at the QCD-94 Workshop,
Montpellier, July 7-13 1994}
\baselineskip=16pt
\end{flushleft}
\newpage

\section{Sum Rules on the Light-Cone}

The standard way to derive QCD sum rules for the transition
form factors between two given ground state hadrons
starts from the vacuum correlation function of three
currents.
Here we demonstrate an alternative method which may be
used in cases where one of the hadrons is a light
meson.

\subsection{$B (D)\ra \pi$ form factors}

An important example is provided by the
form factor
$f^{+}_{\pi}(p^2)$ which enters the matrix element
$<\pi(q)|\bar{u} \gamma_\mu b |B(p+q)>$, the momenta
given in brackets. Consider the correlation function
\be
<\pi(q)\mid T\{\bar{u}(x)\gamma_\mu b(x),
\bar{b}(0)i\gamma_5 d(0)\}\mid 0>
\label{1}
\ee
between the vacuum and an on-shell pion state.
It is possible to
calculate this correlation function in the
kinematical region of highly virtual b-quarks.
After expanding the $b$-quark propagator near the light-cone,
the correlation function (\ref{1}) can be expressed in terms of
light-cone wave functions of the pion,
i.e. nonlocal matrix elements of the type
$< \pi(q) \mid \bar{u}(x) \Gamma_a d(0) \mid 0 >$
where $ \Gamma_a$ denotes a combination of Dirac matrices.
Over the years a great deal has been learned about these wave
functions (see e.g \cite{CZ,BF}).
All low-twist wave functions have been
identified and their asymptotic form has been determined.
Also the nonasymptotic corrections have been estimated.

In order to extract the desired form factor from the
correlation function (\ref{1})
we employ a QCD sum rule with respect to the $B$-meson channel.
After Borel transformation
one arrives at an expression of the form:
\bq
f^+_\pi( p^2)= \frac{f_\pi m_b^2}{2f_Bm_B^2}\int_
\Delta^1\frac{du}{u}(\varphi_\pi(u) + ...)
\nonumber\\
exp \left(\frac{m_B^2}{M^2}
-\frac{m_b^2-p^2(1-u)}{uM^2}\right) ~.
\label{formf}
\eq
Here $\varphi_\pi(u)$ represents the leading twist 2
pion wave function, while the ellipses denote contributions
of higher twists. $M^2$ is the Borel parameter and
$f_B$ is the $B$ meson decay constant as
determined from an appropriate two-point sum rule.
The integration limit $\Delta = (m_b^2-p^2)/(s_0-p^2) $
depends on the effective threshold $s_0$ above which the
contribution from higher states in the $B$-channel is subtracted.
In our calculation of $f^+_\pi$ we have included
quark-antiquark wave functions up to twist four. In addition, we
have also evaluated the contributions from
quark-antiquark-gluon wave functions associated with the
first-order correction to the free $b$-quark propagation.
Further details can be found in \cite{BKR}.

In Fig. 1 numerical results  are plotted for $f^+_\pi$.
Making obvious replacements in (\ref{1}) and (\ref{formf}),
one can easily obtain the corresponding sum rule
for the $D \ra \pi$ (and also $ B\ra K$) form factor.

\subsection{$B^*B\pi$ ($D^*D\pi$) couplings}
The same correlation
function (\ref{1})
considered above can also serve as a starting point for a calculation
of the $B^*B\pi$ coupling \cite{BBKR} defined by
the hadronic matrix element
\be
\langle B^{*0}(p)\pi^+(q)\mid B^+(p+q)\rangle =
-g_{B^*B\pi}q_\mu \epsilon ^\mu .
\label{g}
\ee
However, in contrast to 1.1 one now has to employ sum rules in both
the $B-$ and $B^*-$
channels carrying the momenta $p+q$ and $p$ , respectively.
After double Borel transformation  one obtains an expression for
the product $f_Bf_{B^*}g_{B^*B\pi}$ the leading twist term of which
depends
on the pion wave function $ \varphi_\pi(u)$ at a fixed momentum fraction
$ u \simeq 1/2 $. Taking $ \varphi_\pi(1/2)=1.2 \pm 0.2 $
together with the corresponding values of the
higher twist wave functions \cite{BF} and dividing by the values of
$f_B$ and $f_{B^*}$ as given by appropriate two-point sum rules,
we find, numerically,
$g_{B^*B\pi}= 29\pm 3 $ and $g_{D^*D\pi}= 12.5\pm 1$.
{}From the latter value we predict the partial width
$ \Gamma (D^{*+} \ra D^0 \pi^+ )=32 \pm 5~ keV $.
This estimate is consistent with the limit derived from recent
measurements \cite{ACCMOR,CLEO1}.
Further details and a comparison with other estimates
are given in \cite{BBKR}.

In conclusion, we emphasize that light-cone sum rules
represent a useful alternative to the
conventional QCD sum rule method.
In this variant, the nonperturbative aspects are
described by a set of
wave functions on the light-cone with different twist and quark-gluon
multiplicity.
These universal functions can be studied in a variety of processes
involving the $\pi$ and $K$ meson, or other light mesons.

\section{ A Sum Rule for  $B \ra J/\psi K $ }

Nonleptonic two-body decays of heavy mesons are usually calculated
in the factorization approximation for the appropriate matrix element of the
weak Hamiltonian. However, as well known, naive factorization fails.
In order to achieve agreement with experiment it is necessary to let the
coefficients $a_{1,2}$ multiplying the matrix elements deviate from their
values $ a_{1,2}= c_{1,2} +c_{2,1}/3 $ predicted in short-distance QCD.
Phenomenologically \cite{BSW}, they are treated as free
parameters to be determined from experiment.

The decay $B \rightarrow J/\psi K $ provides an important example.
In factorization approximation the amplitude
is proportional to $a_2f^+_K f_{\psi}$ where $f^+_K$ is the
$B \ra K $ form factor at $p^2=m_\psi^2$ and $f_{\psi}$ is
the $J/\psi$ decay constant. From the short-distance value for $a_2$,
the branching ratio is estimated to be almost an order of magnitude
smaller than the experimental value \cite{CLEO}. On the other hand,
dropping the term proportional to $c_1/3$ in $a_2$
as suggested by $1/N_c$ expansion \cite{BGR}
yields reasonable agreement. In fact, the factorizable term proportional
to $c_1/3$ may be cancelled by nonfactorizable contributions being of the
same order in $1/N_c$.
Such a cancellation was first advocated in
\cite{BGR} and then shown in \cite{BS} to actually
take place in two-body $D$-decays using QCD sum rule techniques
in order to estimate the nonfactorizable
amplitudes.

Recently, we have investigated
the problem of factorization in $B$ decays using
$B \rightarrow J/\psi K $ as a study case \cite{KLR}.
Following the general idea put forward in \cite{BS},
we calculate the four-point correlation function
\be
<0\mid T\{j_{\mu5}^K(x)j_\nu^\psi(y)H_W(z)j^B_5(0)\}\mid 0>
\label{corr}
\ee
by means of the short-distance OPE.
Here $ j_{\mu5}^K= \bar{u}\gamma_\mu \gamma_5s $ ,
$ j_\nu^\psi= \bar{c}\gamma_\nu c $ and
$ j^B_5= \bar{b}i\gamma_5 u $
are the generating currents of the mesons involved and
$H_W$ is the effective weak Hamiltonian.
To the lowest nonvanishing order in $\alpha_s$ the
nonfactorizable contributions to (\ref{corr}) only arise from the
operator
$\tilde{O_2}=(\bar{c}\Gamma^\rho\lambda^ac)
(\bar{s}\Gamma_\rho \lambda^ab)/4$, where
$\Gamma_\rho = \gamma_\rho(1-\gamma_5)$. Obviously, the contribution of this
operator to the $B \rightarrow J/\psi K $
amplitude vanishes by factorization.
Parametrizing the nonfactorizable
matrix element by $\langle J/\psi K \mid \tilde{O}_2 \mid B\rangle =
2\tilde{f}f_\psi(\epsilon^\psi \cdot q)$ we construct a sum rule
for $\tilde{f}$ which enters the correlation function
(\ref{corr}) through the ground state
contribution. In the QCD part of this sum rule all nonperturbative
contributions to (\ref{corr}) from vacuum condensates up
to dimension 6 are included.
In the hadronic part a complication arises from
intermediate states in the $B-$meson channel carrying the quantum
numbers of a $\bar{D}D_s^*$ pair. These virtual states are
created by weak interaction and converted into the
$J/\psi K $ final state by strong interaction.
By examination of the quark-gluon representation of
(\ref{corr}) one can identify corresponding four-quark
$\overline{u}s\overline{c}c$ intermediate states. Invoking quark-hadron
duality we cancel
this piece of the QCD part against the unwanted hadronic
contribution.

We then perform, as usual, a
Borel transformation in the $B-$meson channel and take
moments in the charmonium channel.
The spacelike momentum squared in the $K$-meson channel is
kept fixed. Since, as explained in \cite{KLR},
subtraction of higher state contributions  using the local quark-hadron duality
is not possible in this case, we take them into account by using a simple
two-resonance description in each of the three channels.
This approximation, finally, yields
$\tilde{f} = -(0.045 \div 0.075)~$.
\label{ftilde}
The full $B\ra J/\psi K $ decay amplitude is proportional to
\be
a_2=c_2+\frac{c_1}3 + 2c_1\tilde{f}/f_K^+
\label{a2}
\ee
where the first two coefficients are associated with the factorizable matrix
element, while the third term represents the leading nonfactorizable
contribution. We find that the factorizable nonleading in $1/N_c$
term and the nonfactorizable term in (\ref{a2})
are opposite in sign. Although the nonfactorizable matrix
element is considerably smaller than the factorizable one,
$|\tilde{f}/f^+_K|\simeq 0.1 $, it has a strong quantitative impact
due to its large coefficient,
$|2c_1/(c_2+c_1/3)| \simeq 20 \div 30 $. In fact, if
$|\tilde{f}|$ is close to the upper end of the predicted
range, the nonfactorizable contribution almost cancels
the factorizable one proportional to $c_1/3$. This is
exactly the scenario anticipated by $1/N_c$-rule \cite{BGR}.

It is very interesting to note that our theoretical estimate yields
a negative overall sign for $a_2$ in contradiction to a global fit
to data \cite{CLEO}.
Furthermore, there is no theoretical reason in our approach
to expect universal values or even
universal signs for the coefficients $a_{1,2}$ in different channels.
Universality can
at most be expected for certain classes of decay modes, such as
$B\rightarrow D\pi$ or $B \rightarrow D\overline{D}$, etc.
Also, there is no simple relation between $B$ and $D$ decays
in our approach since the  OPE for the corresponding correlation
functions significantly differ in the relevant diagrams and in the
hierarchy of mass scales.
We hope to be able to clarify these issues further.

Concluding we would like to stress that QCD
seems to predict a much richer pattern in two-body weak
decays than what is revealed by the current phenomenological
analysis of the data.

\vspace{1cm}
{\bf Acknowledgements}.

R.R. thanks Stephan Narison for
organizing this fruitful workshop.
This work is supported by the German Federal Ministry for Research and
Technology (BMFT) under contract No. 05 6MU93P.

\newpage
\begin{figure}[1]
\centerline{
\begin{picture}(500,500)(0,0)
\put(0,0){\strut\epsffile{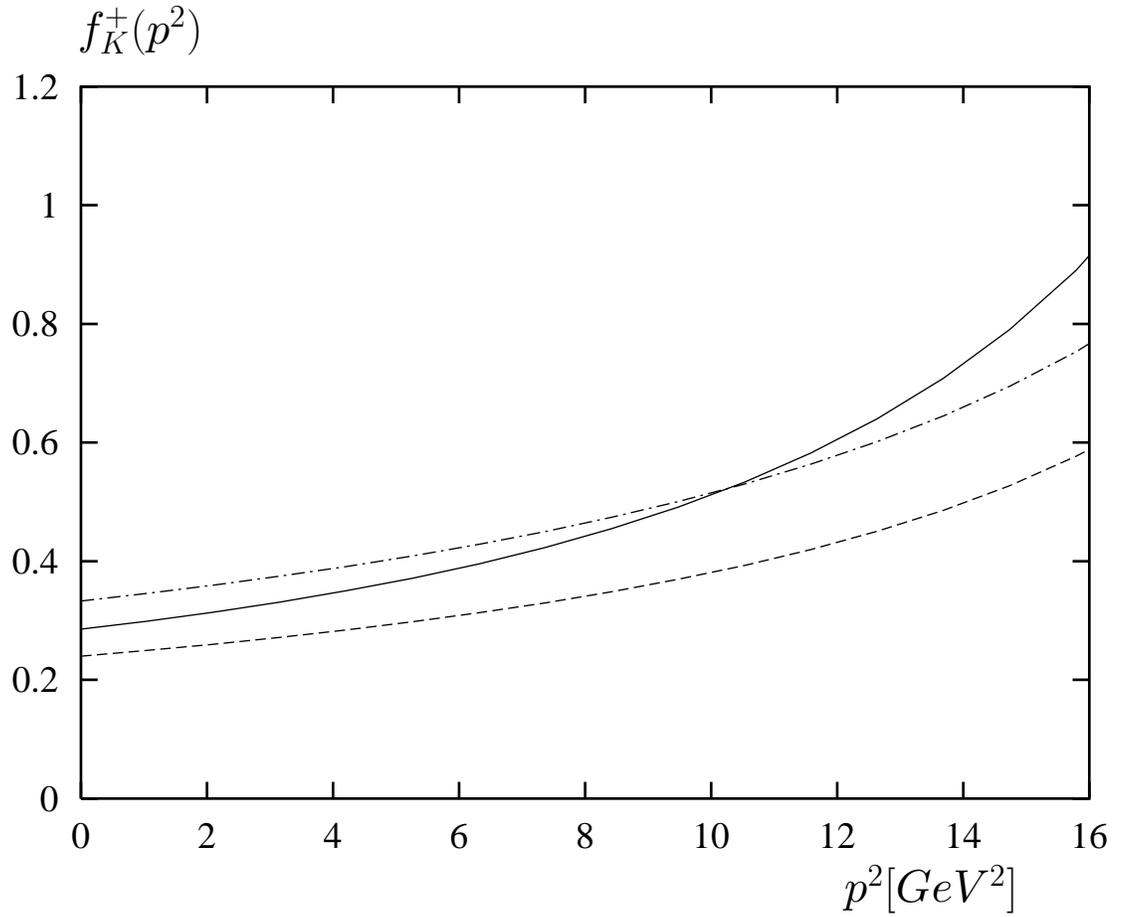}}
\put(100,335){{\Large $f^+_{K}(p^2)$}}
\put(390,10){{\Large $p^2[GeV^2]$}}
\end{picture}
}
\caption[dummy]
{ $B \ra \pi $ form factor obtained from the light-cone sum rule
(solid curve). The quark model prediction [7]
(dash-dotted curve) and the result of a conventional
three-point sum rule [8] (dashed curve) are shown for comparison.
}
\end{figure}

\newpage

\begin{thebibliography}{20}

\bibitem{CZ} V.L. Chernyak, A.R. Zhitnitsky,
Phys. Rep. 112 (1984) 173.

\bibitem{BF} V.M. Braun, I.B. Filyanov,
 Z. Phys. C44 (1989) 157; ibid.
C48 (1990) 239.

\bibitem{BKR} V.M. Belyaev, A. Khodjamirian, R. R\"uckl,
 Z. Phys.  C60 (1993) 349.

\bibitem{BBKR} V.M. Belyaev, V.M. Braun, A. Khodjamirian,
R. R\"uckl, preprint MPI-PhT/94-62.

\bibitem{ACCMOR}
ACCMOR Collab., S. Barlag et al., Phys. Lett. B278 (1992) 480.
\bibitem{CLEO1}
CLEO Collab., F. Butler et al., Phys. Rev. Lett. 69  (1992) 2041.
\bibitem{BSW} M. Bauer, B. Stech, M. Wirbel, Z. Phys.
C34 (1987) 103.
\bibitem{BBD}  P. Ball, V. Braun, H. Dosch, Phys. Lett.
 273B (1991) 316 .
\bibitem{CLEO}
M.S. Alam et al. , CLEO preprint CLNS 94-1270 (1994).
\bibitem{BGR} A.J. Buras, J.-M. Gerard, R. R\"uckl,
Nucl. Phys. B268 (1986) 16.
\bibitem{BS} B.Yu. Blok, M.A. Shifman, Sov. J. Nucl.
Phys.  45 (1987) 135, 301, 522.
\bibitem{KLR} A. Khodjamirian, B. Lampe, R. R\"uckl, {\it in
preparation}.
\end{thebibliography}
\end{document}